\documentclass[iop]{emulateapj-rtx4} 
\shortauthors{Sekanina}
\shorttitle{A New Kreutz Sungrazer C/2024 S1 (ATLAS)}
\slugcomment{Version \today }

\begin{document}
\title{Comet ATLAS (C/2024 S1) --- Second Ground-Based Discovery of\\a Kreutz
 Sungrazer in Thirteen Years\\[-1.6cm]}
\author{Zdenek Sekanina}
\affil{La Canada Flintridge, California 91011, U.S.A.; {\sl ZdenSek@gmail.com}}
\begin{abstract} 
Comet ATLAS (C/2024~S1) is a bright dwarf sungrazer, the second Kreutz
comet discovery~from~the ground this century, 13~years after comet
Lovejoy (C/2011~W3).  The Population~II membership~of comet ATLAS sets
it apart from the overwhelming majority of other bright dwarf sungrazers,
most~of them classified as members of Populations~I, Pe, or Pre-I in the
context of the contact-binary model.  The new sungrazer might be closely
related to comet du Toit (C/1945~X1), but most exciting is the possibility
that it is a fragment of the parent comet of the Great September Comet of
1882 (C/1882~R1) and comet Ikeya-Seki (C/1965~S1).  However, this scenario
requires that the original orbital period of comet ATLAS --- rather poorly
known at present --- be 886~yr.  If its orbital period should turn out to
be decidedly shorter, another scenario involving a 13th-century sungrazer
should be preferred instead.  More work on the orbit needs to be done.
The apparent contradiction between the discovery of comet ATLAS and
previous failures to find any dwarf Kreutz sungrazers in images taken with
large ground-based telescopes at moderate heliocentric distances is
explained by the propensity of Population~II dwarf comets for outbursts,
acting in collusion with extremely rare occurrences of these objects. 
Also addressed are the dynamical properties of perihelion fragmentation
as well as the nature and timing of the expected 21st-century cluster of
Kreutz comets, swarms of dwarf sungrazers,~and~related~issues.

\end{abstract}
\keywords{some comets:\ C/1882\,R1, C/1965\,S1, C/2011\,W3, C/2024\,S1;
methods:\ data analysis{\vspace{-0.16cm}}}

\section{Introduction} 
The diffuse object of magnitude 15 that R.~Siverd detected in CCD images
taken with the 50-cm f/2 Schmidt reflector of the {\it Asteroid
Terrestrial-Impact Last Alert System\/} (ATLAS) project on Mt.~Haleakala,
Hawaii, on September~27, 2024 (Green 2024) turned out to be the second
ground-based discovery of a Kreutz sungrazer in the 21st century, only
13 years after comet Lovejoy (C/2011~W3).  The comet, designated C/2024 S1,
may have been in outburst; it experienced a few more before it has
essentially run out of its pool of volatiles.

When it entered the field of the C3 coronagraph on board the Solar
and Heliospheric Observatory (SOHO) on October~26.5~UT, comet ATLAS
was of magnitude~7 according to the{\vspace{-0.05cm}} {\it Comet
Observation Data\/} (COBS) website.\footnote{See {\tt
https://cobs.si/obs\_list?id=2574}.}  It began to saturate the C3
coronagraph's sensor on October~27.90 UT and the records show that
the brightness first peaked at magnitude 3.4 on 27.94.\footnote{This
is inconsistent with Knight et al.'s (2010) Table~2 that shows C3 images
to begin to saturate at magnitudes between 2.9 and 3.3.}  After a drop
to magnitude~4.3 on 28.10, the comet brightened --- in a fashion
reminiscent of a terminal outburst --- to the peak magnitude of 2.5 on
28.17, 0.32~day before perihelion, after having entered the field of
the C2 coronagraph on 28.12.  Consistent with the peak brightness, the
C2 sensor never saturated. The comet then faded rapidly, its final
traces seen approximately 0.13~day, or 3~hr, before perihelion at
a heliocentric distance of 5\,$R_\odot$.  On the basis of these
observations {\it comet ATLAS is to be categorized as a bright dwarf
Kreutz sungrazer,}\footnote{A dwarf Kreutz sungrazer is defined as
a Kreutz sungrazer that fails to survive its perihelion passage,
disintegrating shortly before.}\,{\it yet~by~no means a typical one\/},
as argued in Section~2.1.
            
\section{$\!$Orbit, Population, and Nature of the Comet} 
More than 300 astrometric observations were available for the orbital
computations by October~23.  Summarized in Table~1, the three independent
sets of elements (JPL 2024, MPC 2024, Nakano 2024) are accurate enough to
leave no doubt that the new Kreutz sungrazer is a member of Population~II:\
the longitude of the ascending node, determined to better than
$\pm$0$^\circ\!$.1, is positioned squarely between those of the Great
September Comet of 1882 (C/1882~R1) and Ikeya-Seki (C/1965~S1), while the
perihelion distance is only a few percent greater than that of either of
these two sungrazers.  However, the orbital arc covered is still too
short to provide an accurate value for the orbital period.  This
uncertainty limits our current knowledge of the comet's history,
an issue discussed~in some detail in Section~3.

\subsection{Relation to Other Kreutz Sungrazers} 
The timing of their perihelion arrivals notwithstanding, no {\it direct\/}
relationship (such as the immediate parent) appears to exist between the
new sungrazer and comet Lovejoy, whose perihelion distance was smaller
by 0.5\,$R_\odot$, whose longitude of the ascending node deviated by
$\sim$20$^\circ$, and which was a member of Population~III.  Undoubtedly,
the two sungrazers have different recent histories.

\begin{table*}[t] 
\vspace{0.15cm}
\hspace{-0.15cm}
\centerline{
\scalebox{1}{
\includegraphics{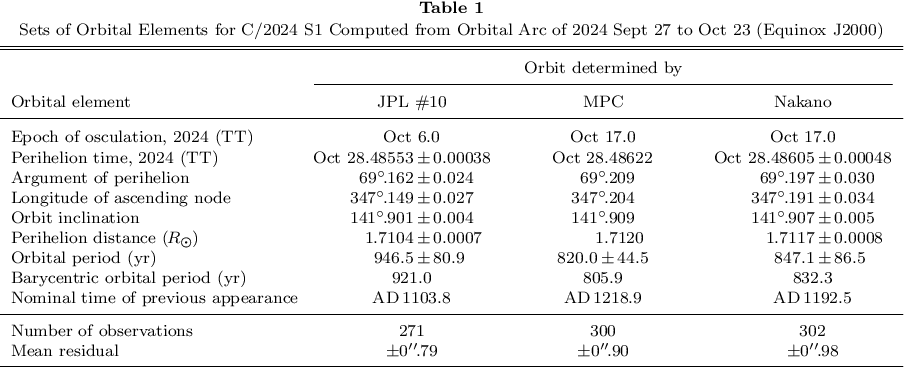}}}
\vspace{0.7cm}
\end{table*}

Studies were conducted in the past 15 years that strove but failed to
detect SOHO-like dwarf Kreutz sungrazers from the ground, using large
telescopes, at moderate solar elongations before the objects entered
the field of the C3 coronagraph (e.g., Knight et al.\ 2010, Ye et al.\
2014).  Discovery of comet ATLAS at magnitude~15 at 1.07~AU from the
Sun looks altogether incomprehensible in the light of the
unsuccessful searches for objects of this kind, reaching{\vspace{0.16cm}}
magnitude~23.

In an effort to understand this disparity, I studied the dwarf Kreutz
sungrazers, observed with the SOHO coronagraphs to be brighter at maximum
light than apparent magnitude~3, as a function of their orbit-dependent
population membership.  Table~2 presents a chronological list of the
28~objects that I came up with, collecting the data from (i)~Knight et
al.'s (2010) photometric study, which covered 1996--2005, especially
their Table~2; and (ii)~Sekanina \& Kracht's (2013) investigation,
which, encompassing 2004--2013, was undertaken with the aim to examine
an apparent swarm of bright dwarf sungrazers in the context of comet
Lovejoy's arrival at the end of 2011.  For each sungrazer in column~1 of
the table the peak magnitude (in the field of the C2 or C3 coronagraph)
is in column~2.  The applied routine, described in Sekanina (2022a),
begins with the set of {\it nominal\/} parabolic elements, computed by
Marsden for most of the tabulated sungrazers or by Kracht for the objects
after mid-2010.  The nominal longitude of the ascending node is corrected
approximately for the{\vspace{-0.06cm}} nongravitational effect, which
establishes the {\it true\/} longitude of the ascending node,
$\widehat{\Omega}$, in column~3, and determines the population
membership, in column~4.

\begin{table}[b] 
\vspace{0.5cm}
\hspace{-0.19cm}
\centerline{
\scalebox{0.965}{
\includegraphics{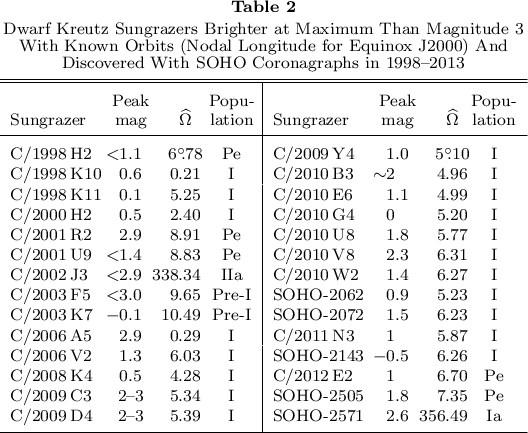}}}
\vspace{0.06cm}
\end{table}

An astonishing feature of Table 2 is the absence of any sungrazers of
Population~II, whose member comet \mbox{ATLAS} is!  An overwhelming
majority (nearly 70~percent) of the tabulated objects comes from
Population~I, which is known to be associated with the Great March
Comet of 1843 (C/1843~D1).  An additional 25~percent of the set
belongs to the related Populations~Pe (associated sungrazer C/1963~R1,
Pereyra) or Pre-I (no naked-eye associated sungrazer known).  Only
single objects in the table appear to be members of Populations~Ia
and IIa, respectively.  It is well-known that Population~I dominates
Population~II among all SOHO sungrazers with known orbits:\ my recent
estimate of the abundance ratio has been 14:1 in the least (Sekanina
2022a), but Table~2 suggests that among the bright SOHO objects the
ratio is still likely to be quite a bit higher.

Nondetection of typical dwarf Kreutz comets at moderate heliocenntric
distances from the ground thus implies that Population~I (and similar)
sungrazers are at those distances much fainter than Population~II
sungrazers and that therefore intermittent outbursts are common among the
latter objects but absent among the former ones:\ the two populations
fundamentally differ from one another in the physical makeup.  This
hypothesis is supported by the observed light curve of comet ATLAS,
which consisted mostly of a sequence of outbursts.

In the era of SOHO, comet ATLAS has been~the~first dwarf Kreutz sungrazer
discovered from the ground.  However, it might not be the first
ever object of this particular category.  It has been suspected for
some time (Sekanina \& Kracht 2015) that comet du Toit (C/1945~X1),
discovered at the Boyden Station in South Africa and quite possibly a
member of Population~II, was a dwarf sungrazer as well.  In particular,
it is reasonable to argue that this comet was in outburst when
discovered and observed over a period of just a few nights before
hopelessly lost.\footnote{Marsden (1967) considered comet du Toit
to be a Subgroup~II member, but since all derived orbits (including
more recent ones, e.g., Marsden 1989, Sekanina \& Kracht 2015)
suffer from large uncertainties because of the limited number of
observations over a very short orbital arc, one cannot rule it
out that the comet is a member of Population~Ia.  It is desirable
that attempts be undertaken to search the online available collection
of Boyden plates for additional images of the comet with the aim to
refine its orbit.} 

\begin{table*} 
\vspace{0.15cm}
\hspace{-0.15cm}
\centerline{
\scalebox{1}{
\includegraphics{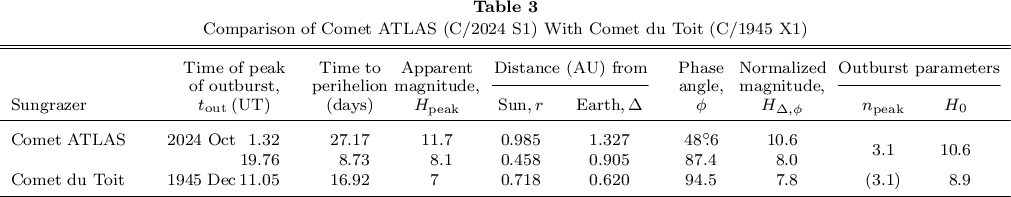}}}
\vspace{0.7cm}
\end{table*}

\subsection{Comparison With Comet du Toit}
On the assumption that comet du Toit was discovered and observed only
during an outburst, the reported apparent magnitude~7 on 1945 December~11
can be used to approximately compare the intrincic brightness of this
object with comet ATLAS.  This allows one to assess the chances that
comet du Toit was indeed a dwarf sungrazer and, if it was, to determine
its place in the hierarchy of these comets.

Let comet ATLAS or another object of its kind experience an outburst,
during which it reaches a peak apparent magnitude $H_{\rm peak}$.  Let
the heliocentric and geocentric distances at the time be, respectively,
$r$ and $\Delta$ and the phase angle $\phi$.  The magnitude normalized
to a unit geocentric distance and to a phase angle 0$^\circ$,
$H_{\Delta,\phi}$, equals
\begin{equation}
H_{\Delta,\phi}(r) = H_{\rm peak} - 5 \log \Delta - f(\phi),
\end{equation}
where $f(\phi)$ is the phase law, that is, the brightness ratio of the
dust coma at phase angles $\phi$ and 0$^\circ$, expressed in magnitudes.
If the comet undergoes more outbursts at different heliocentric
distances $r$, I assume that the brightness at their peaks satisfies
a power law $r^{-n_{\rm peak}}$, so that the normalized magnitude
varies as
\begin{equation}
H_{\Delta,\phi}(r) = H_0 + 2.5 \, n_{\rm peak} \log r,
\end{equation}
where $n_{\rm peak}$ is a comet independent constant and $H_0$ is the
peak magnitude for a comet's outburst occurring at a unit heliocentric
distance.  On the assumption that the outbursts of two or more dwarf
sungrazers have amplitudes of comparable magnitudes, $H_0$ measures
the relative sizes of the objects.

I now apply this simple routine to comets ATLAS and du Toit.  For the
phase law I use the formula by Marcus (2007) with the constants for
dust-poor comets, because the observations of comet ATLAS suggest an
overwhelming contribution from C$_2$ emission over dust to the light of
the coma during the outbursts.  Presented in Table~3 are the data for
the two most obvious flare-up events in this comet's light curve, one
peaking on October~1, the other 18~days later.  The peak apparent
magnitudes are the visual estimates from the COBS database.

For comet du Toit I employ the data from the original report by
Paraskevopoulos (1945).  The outburst time is equated with the
discovery time and the estimated magnitude 7 is obviously photographic.
As the magnitudes for comet ATLAS are visual, one should keep in mind
application of an unknown correction because of the color index.  The
comparison in Table~3 shows that du Toit is likely to be brighter than
comet ATLAS.  The color index increases the nominal difference of
1.7~magnitude by about 0.5, although this may be the uncertainty with
which the difference is determined in the first place.  Given that
comet ATLAS was magnitude $\sim$3 at maximum light, one can crudely
estimate that comet du~Toit may have reached a peak magnitude of
perhaps +1, which --- as is apparent from Table~2 --- still makes it
a dwarf sungrazer that fails to survive perihelion.

Another suspicious circumstance is the difference of 79~years between
the perihelion arrivals of the two comets, an issue that may be linked
to the general fragmentation pattern of the Population~II sungrazers
(Section~4).  Major progress and more definitive conclusions on comet
du~Toit are predicated on a successful search for the object's
additional images in a collection of Boyden plates recently digitized
by the Harvard College Observatory's DASCH Project leading to an
improved orbit.

\section{Possible History}
The troublesome disparity and uncertainty of the data on the orbital
period in Table~1 rule out any definitive conclusions on the history
of comet ATLAS at this time.  Indeed, accepting conservatively an
error of $\pm$3$\sigma$, the comet could have previously passed
perihelion anytime between about AD~850 and 1450.  A large number
of potential Kreutz sungrazers was reported from this period of
time, starting with a Japanese-Chinese comet of 852 and ending with
a Japanese comet of 1434 (Hasegawa \& Nakano 2001, England 2002).

Very exciting is the emerging possibility that comet ATLAS was another
fragment of the parent to the sungrazers of 1882 and 1965, a scenario that
could turn out (i)~to provide a new test for the parent's identity, in an
unlikely case that the uncertainty of the true orbital period of the new
comet could eventually be brought down to not more than several years;
and, if all three objects are indeed shown to have common origin, (ii)~to
reveal the remarkable similarity of the fragmentation sequences of the
parent comet and the 1882 sungrazer, its primary product.

The relationship, addressed in detail below, requires that the
orbital period of comet ATLAS be near 900~yr.  If it is much shorter,
the sungrazer is instead a fragment of another Kreutz sungrazer.  A
likely candidate for the parent is then the comet of 1232, in which case
the orbital period of 792~yr is needed, within 14~yr (0.31$\sigma$)
of the nominal value for comet ATLAS predicted by the MPC orbit and
within 40~yr (0.46$\sigma$) of the value predicted by Nakano's
orbit.  According to Hasegawa \& Nakano (2001) the comet of 1232 was
observed in both China and Japan between October~17 and early December
of that year.  Its motion was consistent with that of a Kreutz comet
passing perihelion on October~14\,$\pm$\,2~days; this comet is also
listed in Ho's (1962) catalogue as No.~429 and mentioned (as a Chinese
comet) by England (2002).  The proposed identity is of course uncertain
also because it is not known whether the comet of 1232, even if a Kreutz
sungrazer, was a member of Population~II.

The scenario that involves comet ATLAS with the sungrazers of 1882 and
1965 enjoys a great advantage in that all three fragments are known
to undoubtedly be members of the same population, with the orbits of
superior quality relative to those in any other scenario.  Marsden
(1967) was the first who tackled the problem of the common origin of
the Great September Comet of 1882 and Ikeya-Seki.  Having adopted
Kreutz's (1888, pp.~108--110) view that C/1106~C1, one of the
most famous historical comets, was the previous appearance of the
sungrazer of 1882, Marsden integrated back in time the orbits of
this comet's principal nucleus and the brighter nucleus of comet
Ikeya-Seki to find that the two comets were one before their 12th
century perihelion.  Indeed, given that the sungrazers of 1882 and
1965 were both observed to fragment, it is logical to expect that
their parent should have done so as well.

Given the apparent orbital period between 700~yr and 800~yr, the
sungrazer of 1882 would be an obvious 19th century candidate for
the next appearance of the 1106 comet, if it were not for an equally
eligible \mbox{competitor}, the bright sungrazer of 1843.  Historically,
however, this latter comet was (mistakenly) suspected to represent the
return of the comet of 1668, as eloquently described elsewhere (e.g.,
Lynn 1882, Seargent 2009), although~both Hubbard (1852) and later
Kreutz (1901) determined that the orbital period of the 1843 sungrazer
was much longer than the implied 175~years; a more recent work showed
that it was comparable to that of the 1882 sungrazer before splitting
(Sekanina \& Chodas 2008).  Under these circumstances, Kreutz's suggestion
of the identity of the 1106 and 1882 comets presents one of two a priori
equally probable hypotheses, the other one linking the comet of 1106 to
the sungrazer of 1843.  The ramifications of the problem of a {\it missing
12th century sungrazer\/} could in fact be formulated even more broadly:\
{\it the pair of brilliant 19th century sungrazers of nearly equal orbital
periods implied the existence of two prominent 12th century sungrazers,
but the historical records appeared to reveal only one\/}.  In the absence
of any other relevant information there was no reason for preferring
either option.

The problem of which is which has recently~come~to~a head when an
in-depth reexamination of the motion~of Ikeya-Seki's brighter nucleus A,
linked with the preperihelion motion of the single (presplit) nucleus,
indicated that this comet must have previously been at perihelion around
AD~1140~and could not have originated in any breakup of the 1106 sungrazer
(Sekanina \& Kracht 2022).  This conclusion happens to correlate
remarkably well with the determination by Marsden (1967) that the 1882
comet had previously been at perihelion in April 1138, a result
he obtained by integrating back in time Kreutz's (1891) best orbit for
the comet's principal (and, presumably, most massive) nucleus B, yet
dismissing the derived time as ``uncertain''!

By default, the comet of 1106 must have returned as the sungrazer of
1843, so the work on Ikeya-Seki took care of one part of the problem.
The other part, the identity of the 12th century parent to the 1882 and
1965 sungrazers, was subsequently resolved when we came across the
historical record of a Chinese comet in Ho's (1962) catalogue under No.~403,
observed~in~September~1138 (Sekanina \& Kracht 2022).\footnote{Ho also
lists a {\it Japanese\/} comet under the same number, which is definitely
independent of the comet recorded by the Chinese.}  In all probability,
this is the {\it missing\/} 12th century sungrazer, even though, moving
about the Sun on the far side from the Earth, it did not look at all
impressive; it reached perihelion around 1138 August~1.  This scenario
is supported by the orbit integration of the sungrazers of 1882 and 1965,
as their perihelion elements are in much better agreement in 1138 (some
within the mean errors) than in 1106.

The exact condition that this scenario prescribes for the true (barycentric)
orbital period of comet ATLAS is \mbox{$P_{\rm orig} = 886.2$ yr}.  At
present, both the JPL and Nakano's orbits are formally consistent with
this scenario, to 35~yr (0.43$\sigma$) and 54~yr (0.62$\sigma$),
respectively, but mainly because of the large uncertainties of the
orbital period.  The MPC orbit in Table~1 is not in line with this
scenario, implying an orbital period that is off by about 1.8$\sigma$.

In a broader context of the contact-binary model~for the Kreutz system
(Sekanina 2021), the previous appear\-ance of the 1138 comet was {\it
Fragment~II\/} in a~\mbox{compact} cluster of the first-generation
fragments\,(Sekanina 2022b) that arrived at perihelion in late AD~363 and
was briefly recorded by the Greek historian Ammianus Marcellinus.  If
{\it Fragment~II\/} broke up at this perihelion into two or more pieces,
the principal fragment (to become the comet of 1138) may have been
accompanied by one to become the comet of 1232.\footnote{In a paper
on the super-Kreutz system (Sekanina 2023) I considered the 1232
comet to be a member of Population~IIa, but this was merely an
example of potential sungrazers' classification.}

Note that comet ATLAS should be a very late fragment (\mbox{$U_{\rm
frg} \gg 0$}) if derived from the comet of 1138, but an early fragment
(\mbox{$U_{\rm frg} < 0$}) if coming from the comet of 1232; either
way, its grandparent is the same.  Further work may hopefully tell
which of the two scenarios is preferable.  I see no other viable
hypothesis at~this~time.\,\,\,

\section{Comparison of Fragmentation Sequences of C/1882~R1 and Its
 Parent} 
In a very recent paper written before the discovery of comet ATLAS I discussed
the motions of the nuclear fragments of the 1882 sungrazer, based on the
assumption that the breakup involved no momentum exchange among the
fragments (Sekanina 2024a).  Their orbital periods are then determined
by the positions, at the time of separation, of their centers of mass
relative to the center of mass, {\it CM\/}, of the presplit nucleus.
If the length of the nucleus along the radius vector at this time is
$\Delta U$, its orbital period is $P_{\rm par}$, and the nucleus splits
at a heliocentric distance $r_{\rm frg}$, a fragment whose center of mass
deviates from {\it CM\/} along the radius{\vspace{-0.065cm}} vector by
$U_{\rm frg}(r_{\rm frg},P_{\rm par})$ \mbox{($-\frac{1}{2}\Delta U \!<\!
U_{\rm frg} \!<\! +\frac{1}{2} \Delta U$} for a symmetric body) ends
up in an orbit whose orbital period $P_{\rm frg}$ is{\vspace{-0.05cm}}
with a relative precision on the order of 10$^{-5}$ given by
\begin{equation}
P_{\rm frg} = P_{\rm par} \! \left[ 1 - \frac{c_0 U_{\rm frg}(r_{\rm frg},
 P_{\rm par})}{r_{\rm frg}^2} P_{\rm par}^{\frac{2}{3}} \right]^{\!
 -\frac{3}{2}} \!\!,
\end{equation}
where $P_{\rm par}$ and $P_{\rm frg}$ are in yr, $r_{\rm frg}$ in AU,
and{\vspace{-0.07cm}} the~constant \mbox{$c_0 =
2\:$AU$\:$yr$^{-\frac{2}{3}}\,$or$\:1.337 \!\times\!
10^{-8}\:\!$AU$^2\:\!$km$^{-1\:\!}$yr$^{-\frac{2}{3}}$}~when~$U_{\rm frg}$
is in AU or km,~respectively.

The orbital periods of the presplit nucleus and its fragments should
strictly be taken at the time of fragmentation, which is at or near the
perihelion point, but is not known exactly.  Of particular interest are
the minimum values of the distances $U_{\rm frg}$, which offer information
on a lower limit of the nucleus dimensions; one obtains from Equation~(3)
in this case
\begin{equation}
U_{\rm min} = \frac{q^2}{c_0} \! \left(\!P_{\rm par}^{-\frac{2}{3}} \!-\!
 P_{\rm frg}^{-\frac{2}{3}} \right) \!,
\end{equation}
where $q$ is the perihelion distance in AU.  Given that $U_{\rm frg}$
varies as the square of $r_{\rm frg}$, it could attain a maximum value
of $4U_{\rm min}$ along the perihelion arc between the true anomalies
of $-$90$^\circ$ and +90$^\circ$.  In line with this general rationale,
the osculating orbital periods of the four nuclear fragments derived
by Kreutz (1891) were next converted into the barycentric future orbital
periods and used in Equation~(4), with nucleus~B (No.~2 in Kreutz's
notation) adopted as the principal (most massive) fragment that moved
essentially in the orbit of the presplit nucleus (with \mbox{$U_{\rm
min} = 0$}).

\begin{table}[t] 
\vspace{0.17cm}
\hspace{-0.17cm}
\centerline{
\scalebox{1}{
\includegraphics{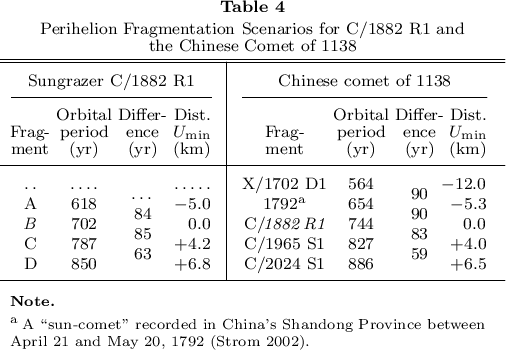}}}
\vspace{0.7cm}
\end{table}

The results, derived for the fragments of the 1882 sungrazer in Sekanina
(2024a), are reproduced in Table~4, which for the sake of comparison
also summarizes information on a hypothetical fragmentation sequence
for the 1138 comet.  The known fragments --- the sungrazer of 1882 and
Ikeya-Seki --- are supplemented with two potential fragments observed
in the 18th century:\ a ``sun-comet'', recorded by the Chinese in 1792
(Strom 2002), and X/1702~D1 (Kreutz 1901, England 2002).  These data
are likewise copied from Sekanina (2024a).  Comet ATLAS is the new
addition to the tabulated list of the {\it potential\/} fragments of
the comet of 1138.

The two fragmentation scenarios in Table 4 do share a common feature,
which is remarkable regardless of whether it is fortuitous or not.
Relative to the principal fragment, typed in italics and assigned
\mbox{$U_{\rm min} = 0$}, the other fragments display distinct and
very similar patterns of their center-of-mass positions $U_{\rm min}$.
Sizewise, the 1882 sungrazer comes out to be in the least 12~km across,
but could be 50~km or even more.  The dimensions of the comet of 1138,
which must obviously be larger, are harder to estimate, because it
is by no means certain that the comets of 1792 and 1702 were its
fragments.  Of course neither is this at present clear for comet
ATLAS.

\section{The 21st Century Sungrazer Cluster} 
Marsden may not have been the first who brought~up the question of a
peculiar temporal distribution~of~bright Kreutz sungrazers, but he was
the first to investigate~in detail their apparent{\vspace{-0.21cm}}
clusters, writing (Marsden 1967):\\

\noindent
{\it The times of the{\vspace{-0.02cm}} appearances~of~the~comets~are~\mbox{distributed}
in a highly nonuniform{\vspace{-0.02cm}} manner.~There~seem~to~be~three distinct clusters
of{\vspace{-0.02cm}} comets --- one~in~the~late~\mbox{seventeenth} century, a second in
the{\vspace{-0.02cm}} nineteenth century,~and~a~third in progress at the present time.
The{\vspace{-0.02cm}} strong concentrations in the 1880s and 1960s are{\vspace{-0.21cm}}
particularly noteworthy.}\\

In retrospect, one can figure out what Marsden's definition of a cluster
was likely to be.  His late seventeenth century cluster appears to have
consisted of C/1689~X1, C/1695~U1, and X/1702~D1, the three potential Kreutz
sungrazers over a period of 13~years.  Furthermore, there was C/1668~E1,
another strong suspect, but its inclusion would extend the cluster over
27~years even without the comet of 1702, which is in conflict with the fact
that he did not count C/1945~X1 as part of the twentieth century cluster; 
this limits the extent to less than 20~years for three consecutive objects.
However, since he wrote his paper before comet White-Ortiz-Bolelli
(C/1970~K1) was discovered, his ``strong concentration in the 1960s''
stood for the discoveries of two bright sungrazers over a period of
26~months.  And while, astonishingly, Marsden anticipated the arrival of
a third sungrazer, the comet of 1970 was in one respect a disappointment
to him.  It did not fit his model of two subgroups, moving instead in an
orbit that deviated from Subgroup~I significantly more than Subgroup~II
in both the angular elements and peri\-helion distance, an early warning
of the things to come.\,\,\, 

To sum it up, Marsden's maximum temporal extent for a cluster of three
ground-based Kreutz sungrazers (or potential sungrazers) surviving
perihelion was less than 20~years.  A more relaxed definition of
a cluster suggested the likely existence of eight of them between
$\sim$1550 and 1970 (Sekanina \& Chodas 2007).  It was this compelling
{\it empirical\/} evidence of a long-term quasi-periodicity --- rather
than any theoretical construct --- that led us to predict, in that
same paper, that ``{\it there are \ldots signs that another cluster
of bright Kreutz \ldots comets is on its way to the Sun in the coming
decades, with the earliest objects expected to arrive perhaps as soon
as several years from now\/}.''  Four years after these lines were
written, comet Lovejoy appeared.  Was it a member of the predicted
21st century cluster?  The answer depends on the definition; Lovejoy
was not a member of the cluster in the Marsden sense (unless two more
naked-eye sungrazers arrive in the next few years), but it could be
a member of a broader, less compact cluster.  Besides, the bulk of
the predicted cluster is to arrive in a general time frame of
2025--2050 (see Figure~4 of Sekanina \& Chodas 2007).  On the other
hand, comet ATLAS could not be a member of any such cluster, because
it is a dwarf sungrazer.  In reference to Section~2.2, the same may
apply to comet du~Toit.

Already in his 1967 paper Marsden concluded that the highly nonuniform
distribution of the perihelion times, including the clusters, ``{\it must
be largely fortuitous \ldots\/}''  This view was again voiced by him in
the second paper (Marsden 1989), in which he was dealing extensively with
the sungrazers discovered in the coronagraphs on board the Solwind and
SolarMax missions in the 1980s.  He wrote that the Kreutz comets ``{\it
may be coming back to the Sun more or less continuously.\/}''

Significantly, while disregarding the clusters of bright Kreutz comets
on a time scale of decades and centuries, Marsden (1989) did point out
that some SolarMax sungrazers seemed to be coming in pairs, with perihelia
separated by only 12~days.  This looked like a tendency to clustering
on a time scale orders of magnitude shorter.  Marsden argued that if
caused by differential nongravitational forces acting over the entire
revolution about the Sun, a gap of 12~days between the arrivals of two
SolarMax comets is equivalent to a relative acceleration of \mbox{$\sim
\! 2 \times\! 10^{-12}$\,m~s$^{-2}$}, many orders of magnitude lower than
inferred for companions of the split comets.  One may add that, equally
improbably, the same gap between the arrival times would be equivalent
to a radial distance of 1~meter between the fragments' centers of mass
at the time of separation at perihelion.  And after SOHO discoveries
revolutionized the way the Kreutz system was viewed, the pairs and small
groups of dwarf sungrazers were sometimes arriving merely minutes apart
(Table~4 of Sekanina 2024b), Equation~(4) implying trivial, completely
meaningless ``gaps'' between the fragments' centers of mass, on the order
of 1~mm!

Explanation of these features in the perihelion-arrival distribution of
dwarf sungrazers in terms of fragmentation at perihelion thus became
unsustainable.  Instead, I proposed that the {\it fragmentation process\/},
starting at perihelion, {\it was continuing in a cascading fashion
throughout the orbit about the Sun\/} (Sekanina 2000).

With this major change in the perception of the Kreutz system's
evolution, the genie got out of the bottle.  Fragmentation continuing
episodically throughout the orbit implied {\it much higher fragmentation
rates\/} in comparison to breakups confined to perihelion, which in turn
implied {\it shorter lifetimes of fragments\/} and demanded a {\it tight
limit on the age\/} of the Kreutz system, in contradiction to the large
number of revolutions needed for the indirect planetary perturbations to
explain the sizable gaps between the orbital elements of the two subgroups.
The classic model for Kreutz sungrazers, based {\it exclusively\/} on
tidal disruption at perihelion, collapsed.

The new perception offered new avenues in the understanding of the
nature of the perturbations of fragments' orbital elements.  Whereas
a parent sungrazer's breakup {\it at perihelion\/} has a considerable
effect on each fragment's orbital period and practically none on its
angular elements and perihelion distance, the opposite is true for such
a breakup {\it far from the Sun\/}:\ the {\it normal\/} component of
the separation velocity on the order of 1~m~s$^{-1}$ exerts a change of
several degrees in the longitude of the ascending node and similar changes
in the other two angular elements; the {\it transverse\/} component
changes the perihelion distance by a nontrivial fraction of the Sun's
radius; and the {\it radial\/} component, with a small contribution
from the transverse component, affects each fragment's time of arrival
at the next perihelion by as much as months.  Hence, fragmentation of
a sungrazer at large heliocentric distance does provide a mechanism for
generating a cluster of fragments at the time of the next perihelion
passage.

This rather complex discussion leads to the conclusion that there are
three categories of the clusters of bright Kreutz sungrazers:\ {\it
genuine\/}, {\it fortuitous\/}, and {\it mixed\/}.  In the context of
the contact-binary model the best example of a {\it genuine\/} cluster
was the arrival in AD~363 of the first-generation fragments of the
progenitor's lobes that broke up centuries before that year's perihelion
(Sekanina 2022b). A {\it fortuitous\/} cluster of fragments
of a particular generation is a product of their orbital-period
perturbations, especially at or near perihelion,  that by chance add
up to nearly the same total:\ Let a massive fragment, $X_j$, separating
from a sungrazer, $Y_j$, passing perihelion at time $T_j$, end up in an
orbit of period $P_j$; and, similarly, let fragment $X_k$, separating from
sungrazer $Y_k$ passing perihelion at time $T_k$, end up in an orbit of
period $P_k$.  The two fragments pass perihelion at times that difffer by
\mbox{$\Delta T_{j,k} = |(T_k \!+\! P_k) - (T_j \!+\! P_j)|$}, \mbox{$k \!
\neq\! j$}.  If \mbox{$\Delta T_{j,k} < 20$ yr}, fragments $X_j$ and
$X_k$ are members of a fortuitous cluster \mbox{$(j,k = 1,2,\ldots,n)$}.
Finally, a {\it mixed\/} cluster involves objects, some of which
clustered genuinely, while others fortuitously.  Specific examples show
the differences quite plainly.

The 19th-century cluster consisted of the Great Southern Comet of 1880
(C/1880~C1), the Great September Comet of 1882, and the Great Southern
Comet of 1887 (C/1887~B1).  In the context of the contact-binary model,
the comets of 1880 and 1887, both members of Population~I, made up a
{\it genuine\/} pair, the differences in their orbital elements
consistent with a breakup of their parent, itself a fragment of the
Great Comet of 1106.  The proposed scenario has the 1880 and 1887
comets separating at $\sim$50~AU from the Sun in AD~1135, 29~yr after
the 1106 comet's perihelion, with a relative velocity of 4.0~m~s$^{-1}$,
mostly in the transverse direction (Table~2 of Sekanina 2021), and
with the motion of the companion (the 1887 comet) subjected to a high
sublimation-driven nongravitational deceleration of $\sim$0.0022 the
Sun's gravitational acceleration.  The third object of the cluster,
the brilliant sungrazer of 1882, a member of Population~II and the
proposed principal fragment of the Chinese comet of 1138 (Sections~3
and 4), made a {\it fortuitous\/} pair with the 1880 comet as well as
with the 1887 comet.  As a whole, the cluster in the 1880s was therefore
a {\it mixed cluster\/}.

Turning to the 20th-century cluster, comets Pereyra in 1963 and
Ikeya-Seki in 1965 made up a {\it fortuitous\/} pair, because their
orbital histories have been dramatically different.  As discussed in
Sekanina \& Kracht (2022), the suspected parent of Pereyra was the
comet of 1041, while Ikeya-Seki is believed to be a fragment of the
comet of 1138 (Sections~3 and 4).  The relations of the sungrazers
Pereyra and Ikeya-Seki to comet White-Ortiz-Bolelli in 1970 are less
clear, because the orbital period of the latter is unknown.  However,
the Population~IIa membership of the 1970 sungrazer suggests that
it was a product of perihelion fragmentation of a parent obviously
different from those of the comets of 1963 and 1965.  The 20th-century
cluster was then a {\it fortuitous\/} cluster.

The tight pairs and groups of the dwarf Kreutz sungrazers are at
first sight a different phenomenon.  To distinguish them from the
clusters of the bright objects, I refer to them as {\it swarms\/}
(Sekanina 2024b).  Most --- if not all --- of them are, however,
products of the same progressive fragmentation process primarily
along the post-aphelion branch of the orbit.  The overwhelming
majority of, or possibly all, swarms are thus of the same nature as
the {\it genuine clusters\/}, and the arriving fragments may or may
not belong to the same population.  And, as is the case with the
clusters, the difference in the perihelion time between two
fragments in a swarm is mainly determined by the radial components
of their separation velocities at breakup.  For example, for an
orbital period of 900~yr and radial velocities that differ by
0.5~m~s$^{-1}$, the fragments arrive 0.01~day apart when they
separated 1.4~yr before perihelion at a heliocentric distance of
7~AU; they arrive 0.1~day apart when they separated about 8~yr
before perihelion at a heliocentric distance of some 22~AU; they
arrive 1~day apart when they separated nearly 43~yr before perihelion
at a heliocentric distance of about 63~AU; and they arrive 10~days
apart when they separated approximately 215~yr before perihelion at
a heliocentric distance of some 150~AU, not too far from aphelion but
still more than 200~yr after the passage through that point.

As a stochastic process that steadily changes the mass and size
distributions of fragment populations of the Kreutz system, cascading
fragmentation is constantly increasing the number of sungrazers and
decreasing their dimensions.  The action is accelerated by precipitous
sublimation of the ice-depleted refractory residues of the fragments
just before they reach perihelion, a mechanism that terminates their
existence, thereby balancing the overall number of objects at any
given time.

Cascading fragmentation is bound to affect the clusters of massive Kreutz
comets, but not necessarily the swarms of dwarf sungrazers.  While
increasing numbers of fragments do augment a chance for the formation
of fortuitous clusters of bright objects, the declining dimensions
work in the opposite direction because increasing numbers of fragments
become dwarf sungrazers.  In the long run, we should see ever fewer
brilliant Kreutz sungrazers but more mediocre ones.  This trend should
also apply to the clusters:\ as the sungrazer generation rank grows,
fortuitous clusters become more blurred, with the perihelion times
more evenly distributed.  Bright sungrazers arriving in the 21st
century should still be mostly fragments of the third generation, but
early fragments of the fourth generation cannot be ruled out in the
mix.

In any case, it is virtually certain that a cluster of fairly bright
sungrazers will appear in this century, although it may not be as
tight as the 19th and 20th century clusters.  One can be sure that
the new cluster will not include a sungrazer as bright as the Great
March Comet of 1843~or the Great September Comet of 1882, and it may or
may not contain a sungrazer as bright as Ikeya-Seki.

The swarms of Kreutz dwarf sungrazers should not be subjected to any
major changes in the coming decades, as the members of Population~I
are expected to continue dominating the incoming stream for another
century or more.  However, the arrival rate should be slowly declining
with time (Sekanina 2024a).

\section{Final Comments and Conclusions} 
The currently available orbits for comet ATLAS, the new Kreutz sungrazer
discovered in late September of 2024, leave no doubt that it is a member
of Population~II and related one way or another to the Great September
Comet of 1882 and Ikeya-Seki.  Based on the contact-binary model for the
Kreutz system, the JPL and Nakano's sets of orbital elements in Table~1
are consistent with the direct relationship of the three objects, that
is, the comet of 1882, Ikeya-Seki, and ATLAS sharing the same parent,
the Chinese comet of 1138.  In the scenario preferred by the MPC orbit
and also consistent with Nakano's orbit (but not with the JPL orbit),
comet ATLAS is a fragment of a sungrazer that arrived at perihelion in
the early 13th century, possibly the comet of 1232.  This object and
the comet of 1138 would then both be fragments --- the latter presumably
the principal one --- of {\it Fragment~II\/}, the main mass of the
progenitor's {\it Lobe~II\/}, as follows from the pedigree chart of
the contact-binary model.

A major effort should be expended to refine the orbital period of
comet ATLAS.  In the absence of post-perihelion astrometry, the
options available for improving the quality of the orbital solution
are rather limited.  Given the relatively large mean residual that
describes the currently available orbital results, near
$\pm$0$^{\prime\prime\!}$.8 or higher, the first corrective action
to test should be the incorporation of the nongravitational terms
into the equations of motion.  This is most desirable, as inspection
of Nakano's (2024) list of residuals shows that 13 of the last
14~astrometric observations were rejected because they left large
negative residuals in right ascension, suggesting a strong
deceleration affecting the sungrazer's orbital motion as early as
October~21, or 7~days before perihelion.

If this measure is inadeqate, the incorporation of the astrometry
of the images taken with the C2 coronagraph may be required.
Consideration could also be given to include the astrometry of
the images taken with the NRL's compact coronagraph (CCOR) on board
the recently launched {\it Geostationary Operational Environmental
Satellite\/} (GOES-19), both sets if/when available.  Inclusion
of data from the terminal phase of this comet's orbital motion
may be in the need of application of special procedures, perhaps
similar to those used in the orbital study of comet ISON, C/2012~S1
(Sekanina \& Kracht 2014, Keane et al.\ 2016).  In any case, much
experimentation will almost certainly have to be expended to obtain
a satisfactory orbital solution, if achievable at all.

In the SOHO era, comet ATLAS turned out to be the first Kreutz dwarf
sungrazer discovered from the ground.  It may not be the first such
sungrazer ever, if the suspicion that comet du~Toit --- observed
briefly in December 1945, and possibly also of Population~II --- was
a dwarf sungrazer happens to be confirmed in the future.

The discovery of comet ATLAS appears to contradict the published studies
that no known dwarf Kreutz sungrazers, discovered with the SOHO coronagraphs,
could at moderate solar elongations be detected with large ground-based
telescopes before they enter the field of view of the SOHO instruments.
I find that this disparity is explained by a major difference in the
physical behavior of Population~I vs Population~II objects, acting in
collusion with a nearly complete absence of Population~II objects
among the brighter dwarf sungrazers.  This imbalance is illustrated
in Table~2 by the statistics from the period of 1998--2013 for the
objects, whose peak brightness exceeded magnitude~3:\ Population~I,
68~percent; other populations similar to Population~I, 25~percent;
Population~II, 0~percent.

The culprit is the propensity of the dwarf sungrazers of Population~II
for continually flaring up at moderate heliocentric distances (say, near
and below 1~AU) on the way to perihelion, which makes them much brighter
than the equally massive dwarf sungrazers of Population~I.  (Indeed, the
light curve of comet ATLAS consists almost entirely of outbursts.)  Yet,
the dominance of Population~I causes that the light curve of a {\it
typical\/} dwarf sungrazer is very steep and the object much too faint
at moderate distances from the Sun to detect even with very powerful
ground-based telescopes.
 
A truly remarkable feature of the orbital-period distribution among
massive fragments is the apparent similarity of their progression in
terms of the separation parameter $U_{\rm min}$ between the observed
sequence of the secondary nuclei of the Great September Comet of 1882
and the sequence of actual and potential products of this sungrazer's
presumed parent, the Chinese comet of 1138.  The lengths $U_{\rm
min}$ for the nuclei A, C, and D relative to the principal nucleus B of
the 1882 comet are closely matched by the lengths $U_{\rm min}$ for,
respectively, the sun-comet of 1792 (a potenial fragment), Ikeya-Seki,
and ATLAS (a potential fragment) relative to the comet of 1882, the
principal fragment of the comet of 1138.  But since comet ATLAS does not
really count because of its dwarf status, there is a chance of the arrival
of a possible major fragment soon.  On the other hand, the presence of
the comet of 1702 in the sequence suggests, for the sake of symmetry,
the possible existence of yet another massive fragment to arrive in
the late 21st century or the early 22nd century.

The final points addressed in this paper are related to the issue
of the expected 21st century cluster of bright Kreutz sungrazers.
Referring to the naked-eye objects, Marsden felt that the Kreutz
comets were returning to perihelion more or less continuously and
that their clusters were ``largely fortuitous.''  I find that in
fact there are three types of bright sungrazers' clusters:\ genuine,
fortuitous, and mixed.  By contrast, dwarf sungrazers arrive from
time to time in swarms (or, more often, pairs), which are products
of the process of cascading fragmentation proceeding throughout
the orbit, including the aphelion arc.  Compared to the classical
paradigm, in which Kreutz sungrazers were breaking up only in close
proximity of perihelion, the fragmentation rates are now much higher,
the fragments' lifetimes much shorter, and the age of the Kreutz
system strongly limited.

Because of the dwarf status of comet ATLAS, it is not a member of the
expected 21st century cluster~of Kreutz sungrazers.~It is questionable
whether comet Lovejoy of 2011/12 will end up as part of a broader
cluster, whose key members should be arriving in the coming decades.
Because of the long-term effects of cascading fragmentation, the future
sungrazers should statistically be ever less spectacular and more
mediocre, even though some of them still of naked-eye rank.  In the
context of the contact-binary model, there is no chance for objects in
the forthcoming cluster to be comparable in brightness to the magnificent
sungrazers of 1843 or 1882, and at best only a 50--50~chance for a few to
be as impressive as comet Ikeya-Seki.\\

\begin{center} 
{\footnotesize REFERENCES}
\end{center}
\vspace{-0.45cm}
\begin{description}
{\footnotesize
%
%
%
\item[\hspace{-0.3cm}]
England, K.\ J.\ 2002, J.\ Brit.\ Astron.\ Assoc., 112, 13
\\[-0.57cm]
%
%
\item[\hspace{-0.3cm}]
Green, D.\ W.\ E.\ 2024, CBET 5453
\\[-0.57cm]
%
%
\item[\hspace{-0.3cm}]
Hasegawa, I., \& Nakano, S.\ 2001, Publ.\ Astron.\ Soc.\ Japan, 53, 931
\\[-0.57cm]
%
%
\item[\hspace{-0.3cm}]
Ho, P.-Y.\ 1962, Vistas Astron., 5, 127
\\[-0.34cm]
\item[\hspace{-0.3cm}]
Hubbard, J.\ S.\ 1852, Astron.\ J., 2, 153
\\[-0.57cm]
\item[\hspace{-0.3cm}]
Jet\,Propulsion\,Laboratory\,2024,\,\mbox{\tt
https$\:\!\!$:$\:\!\!$/$\!$/ssd.$\:\!\!$jpl.nasa.gov$\:\!\!$/tools/}{\linebreak}
 {\hspace*{-0.6cm}}{\tt sbdb\_lookup.html\#}
\\[-0.57cm]
\item[\hspace{-0.3cm}]
Keane, J.\ V., Milam, S.\ N., Coulson, I.\ M., et al.\ 2016, Astrophys.{\linebreak}
 {\hspace*{-0.6cm}}J., 831, 207
\\[-0.57cm]
\item[\hspace{-0.3cm}]
Knight,\,M.\,M., A'Hearn,\,M.\,F., Biesecker,\,D.\,A., et 
 al.\,2010,~Astron.{\linebreak}
 {\hspace*{-0.6cm}}J., 139, 926
\\[-0.57cm]
\item[\hspace{-0.3cm}]
Kreutz, H.\ 1888, Publ.\ Sternw.\ Kiel, No.\ 3
\\[-0.57cm]
\item[\hspace{-0.3cm}]
Kreutz, H.\ 1891, Publ.\ Sternw.\ Kiel, No.\ 6
\\[-0.57cm]
\item[\hspace{-0.3cm}]
Kreutz, H.\ 1901, Astron. Abhandl., 1, 1
\\[-0.57cm]
\item[\hspace{-0.3cm}]
Lynn, W.\ T.\ 1882, Obs., 5, 329
\\[-0.57cm]
\item[\hspace{-0.3cm}]
Marcus, J.\ N.\ 2007, Int.\  Comet Quart., 29, 39
\\[-0.57cm]
\item[\hspace{-0.3cm}]
Marsden, B.\ G.\ 1967, Astron.\ J., 72, 1170
\\[-0.57cm]
\item[\hspace{-0.3cm}]
Marsden, B.\ G.\ 1989, Astron.\ J., 98, 2306
\\[-0.57cm]
\item[\hspace{-0.3cm}]
Minor Planet Center 2024, MPEC 2024-V117
\\[-0.57cm]
\item[\hspace{-0.3cm}]
Nakano, S.\ 2024, NK 5323
\\[-0.57cm]
\item[\hspace{-0.3cm}]
Paraskevopoulos, J.\ S.\ 1945, IAUC 1024
\\[-0.57cm]
\item[\hspace{-0.3cm}]
Seargent, D.\ 2009, The Greatest Comets in History:\ Broom Stars{\linebreak}
 {\hspace*{-0.6cm}}and Celestial Scimitars.  New York:\ Springer
 Science+Business{\linebreak}
 {\hspace*{-0.6cm}}Media, LLC, 260pp
\\[-0.57cm]
\item[\hspace{-0.3cm}]
Sekanina, Z.\ 2000, Astrophys.\ J., 542, L147
\\[-0.57cm]
\item[\hspace{-0.3cm}]
Sekanina, Z.\ 2021, eprint arXiv:2109.01297
\\[-0.57cm]
\item[\hspace{-0.3cm}]
Sekanina, Z.\ 2022a, eprint arXiv:2212.11919
\\[-0.57cm]
%
%
\item[\hspace{-0.3cm}]
Sekanina, Z.\ 2022b, eprint arXiv:2202.01164
\\[-0.57cm]
\item[\hspace{-0.3cm}]
Sekanina, Z.\ 2023, eprint arXiv:2305.08792
\\[-0.57cm]
\item[\hspace{-0.3cm}]
Sekanina, Z.\ 2024a, eprint arXiv:2404.00887
\\[-0.57cm]
\item[\hspace{-0.3cm}]
Sekanina, Z.\ 2024b, eprint arXiv:2401.00845
\\[-0.57cm]
\item[\hspace{-0.3cm}]
Sekanina, Z., \& Chodas, P.\ W.\ 2007, Astrophys.\ J., 663, 657
\\[-0.57cm]
\item[\hspace{-0.3cm}]
Sekanina, Z., \& Chodas, P.\ W.\ 2008, Astrophys.\ J., 687, 1415
\\[-0.57cm]
\item[\hspace{-0.3cm}]
Sekanina, Z., \& Kracht, R.\ 2013, Astrophys.\ J., 778, 24
\\[-0.57cm]
\item[\hspace{-0.3cm}]
Sekanina, Z., \& Kracht, R.\ 2014, eprint arXiv:1404.5968
\\[-0.57cm]
\item[\hspace{-0.3cm}]
Sekanina, Z., \& Kracht, R.\ 2015, Astrophys.\ J., 815, 52
\\[-0.57cm]
\item[\hspace{-0.3cm}]
Sekanina, Z., \& Kracht, R.\ 2022, eprint arXiv:2206.10827
\\[-0.57cm]
\item[\hspace{-0.3cm}]
Strom, R.\ 2002, Astron.\ Astrophys., 387, L17
\\[-0.65cm]
%
%
\item[\hspace{-0.3cm}]
Ye, Q.-Z., Hui, M.-T., Kracht, R., \& Wiegert, P.\ A.\ 2014, Astro-\\[-0.09cm]
 {\hspace*{-0.6cm}}phys.\ J., 796, 83}
%
\vspace{-0.41cm}
\end{description}

\end{document}